\documentclass[twocolumn,amssymb, nobibnotes, floatfix, aps, prl]{revtex4-2}
\usepackage[bookmarks=false]{hyperref}
\usepackage[utf8]{inputenc}
\usepackage[english]{babel}
\usepackage{amssymb}
\usepackage{amsmath}
\usepackage{graphicx}
\usepackage{color}
\usepackage{physics}
\usepackage{float}
\renewcommand{\vec}[1]{\boldsymbol{#1}}
\graphicspath{{Figs/}}
\newcommand{\be}{\begin{equation}}
\newcommand{\ee}{\end{equation}}
\newcommand{\bea}{\begin{eqnarray}}
\newcommand{\eea}{\end{eqnarray}}

\def\pref{\eqref}

\begin{document}

\title{Unveiling the pairing Symmetry of the superconducting Sn/Si(111) via angle-resolved THz pump spectroscopy}

\author{Mattia Iannetti$^1$}
\author{Tommaso Cea$^1$}
\author{Cesare Tresca$^{2}$}
\author{Lara Benfatto$^3$}
\author{Gianni Profeta$^{1,2}$}
\affiliation{
$^1$Dipartimento di Scienze Fisiche e Chimiche, Universit\'a degli Studi dell’Aquila, Via Vetoio 10, I-67100 L’Aquila, Italy\\
$^2$CNR-SPIN C/o Dipartimento di Scienze Fisiche e Chimiche, Universit\'a Degli Studi dell'Aquila, Via Vetoio 10, I-67100 L'Aquila,  Italy\\
$^3$Department of Physics, Sapienza University of Rome, P.le A. Moro 2, 00185 Rome, Italy
}

\begin{abstract}
Doping tin surface epitaxially grown on silicon, Sn/Si(111), with boron atoms yields the appearance of a superconducting (SC) phase below $T_c\sim 4-5$K. Even though the pairing mechanism remains unknown, experimental evidence of chiral $d-$wave superconductivity has been recently reported, then ruling out a phonon-mediated pairing.
Here we study theoretically the SC phase and symmetries of the doped Sn/Si(111) within a $t-J$ model.
We analyze the photo-excitation of the system by intense THz pulses and show that the polarization dependence of the induced current can distinguish between different symmetries of the SC gap, thus providing a novel experimental mean to investigate the spectroscopic features of the Sn/Si(111) across the SC transition. 
\end{abstract}

\maketitle

\section{Introduction}

The $\alpha$ phase of monolayer group-IV elements
epitaxially grown on semiconducting surfaces
provides a surprisingly rich platform to study the competition between strongly correlated phenomena,
spin transport and lattice reconstruction.
1/3 monolayer coverage of heavy atoms, like Sn and Pb, 
displays an isoelectronic $\sqrt{3}\times\sqrt{3}R30^\circ$
reconstruction, where the host atoms, the adatoms, occupy the $T_4$ site in an triangular lattice\cite{Carpinelli_nature96},
which is called the $\alpha$ phase.
The three dangling bonds at the surface of the substrate saturate with the adatoms,
leaving a free electron at each $T_4$ site to form a half-filled surface band within the substrate's band gap.
Such a surface band gets strongly affected by the electronic interactions.
Some materials, like Pb/Ge(111), Sn/Ge(111) and Pb/Si(111) exhibit
a low temperature $3\times3$ lattice reconstruction, yielding
a charge-density-wave (CDW), which is often metallic\cite{Santoro_prb99,Cortes_prb13,Hansmann_prl13,
Badrtdinov_prb16,Tresca_prl18,Adler_prl19,Tresca_prb21,Tresca_prb23}.
In contrast, Sn/Si(111) displays a low temperature
Mott insulating phase\cite{Santoro_prb99,
Profeta_prl2007,Hansmann_prl13},
 with the possible development of a $2\sqrt{3}\times\sqrt{3}$ collinear antiferromagnetic (AFM) phase\cite{Li_natcomm13,Lee_prb14}, not yet clearly observed experimentally.
A renewed interest in the study of these systems has been recently provided by the discovery of superconductivity
in the Sn/Si(111) hole-doped with boron\cite{Wu_prl20},
with critical temperatures of the order of: $T_c\sim4-5$K.
Even though the pairing mechanism is still an open question,
it is believed that Sn/Si(111) displays unconventional superconductivity
of electronic origin,
driven by the non-local Coulomb interactions\cite{Wolf_prl22,Biderang_prb22}.
The Ref. \cite{Ming_nature23} recently
found experimental evidence of chiral $d-$wave superconductivity,
ruling out i) a conventional phonon-mediated pairing and
ii) the presence of a spin-triplet superconducting (SC) order parameter (OP).
The present state of the art challenges a deeper understanding
of the nature of the SC pairing observed in the Sn/Si(111)
and the engineering of new experimental protocols to determine the symmetry of the SC gap.
In the last years, the use of intense THz light pulses either in pump-probe protocols or for high-harmonic generation offered a powerful tool to address the physics of several SC systems\cite{Nicoletti_advphot16,Shimano_annrevcmphys20,shovon_natmatrev23}.
Exploiting the non-linear light-matter interaction, this technique allows one to resonantly excite the low energy modes of the superconductor and track their dynamics in the time domain.
The lowest-order non-linear effect induced by the incident light is the third harmonic generation (THG),
that can be strongly enhanced in the SC phase when the frequency of the light lies in the neighboring of the SC gap
$\Delta$\cite{Matsunaga_science14,Matsunaga_prb17,Katsumi_prl18,wang_natmat2018,Giorgianni_natphys19,wang_natphot2019,Chu_natcomm20,shimano_prb20,shimano_commphys21,wang_mgb2_prb21,wangNL_NbN_prb22,wang_mgb2_prb22,wang_natphys22,kaiser_natcomm23,kaiser_cm23}. In addition, the measured dependence of the non-linear response on the light polarization with respect to the main crystallographic axes\cite{Matsunaga_prb17,Katsumi_prl18,shimano_prb20,Chu_natcomm20} can provide additional information on the form of the pairing interaction\cite{Cea_prb18}, on the disorder level\cite{seibold_prb21,Udina_fardisc22,Benfatto_prb23} and on the symmetry of the SC OP\cite{manske_natcomm20}.
So far, non-linear THz photo-excitation has been intensively used to probe both conventional BCS superconductors, like NbN\cite{Matsunaga_science14,Matsunaga_prb17,wangNL_NbN_prb22}, Nb$_3$Sn\cite{wang_natmat2018,wang_natphot2019} and MgB$_2$\cite{Giorgianni_natphys19,wang_mgb2_prb21,wang_mgb2_prb22}, and unconventional high$-T_c$ cuprate\cite{Katsumi_prl18,Chu_natcomm20,shimano_prb23,kaiser_natcomm23,kaiser_cm23} and pnictide\cite{shimano_commphys21,wang_natphys22} superconductors.

In this letter we study theoretically the superconductivity in the doped Sn/Si(111),
assuming that the interactions responsible for the pairing are of AFM nature. 
At the meanfield level, we find two almost degenerate ground states (GS's),
one with chiral $d_{x^2-y^2}+id_{xy}$-wave symmetry and the other with pure $d$-wave,
the latter breaking the $C_3$ symmetry out of $C_6$.
By computing the non-linear current induced by a THz light pulse
as a function of the polarization of the incident light,
we show that the THG is $\pi/3$-periodic for the $d_{x^2-y^2}+id_{xy}$-wave and
$\pi$-periodic for the pure $d$-wave.
Thus, our work proposes the angle-resolved THz pump spectroscopy as an experimental tool to unveil the symmetry of the SC OP in the Sn/Si(111),
alternatively to other spectroscopic techniques based on the quasi-particle interference\cite{Levitan_prb23}.

\section{The model for superconductivity}
\begin{figure}
\centering
\includegraphics[width=\columnwidth]{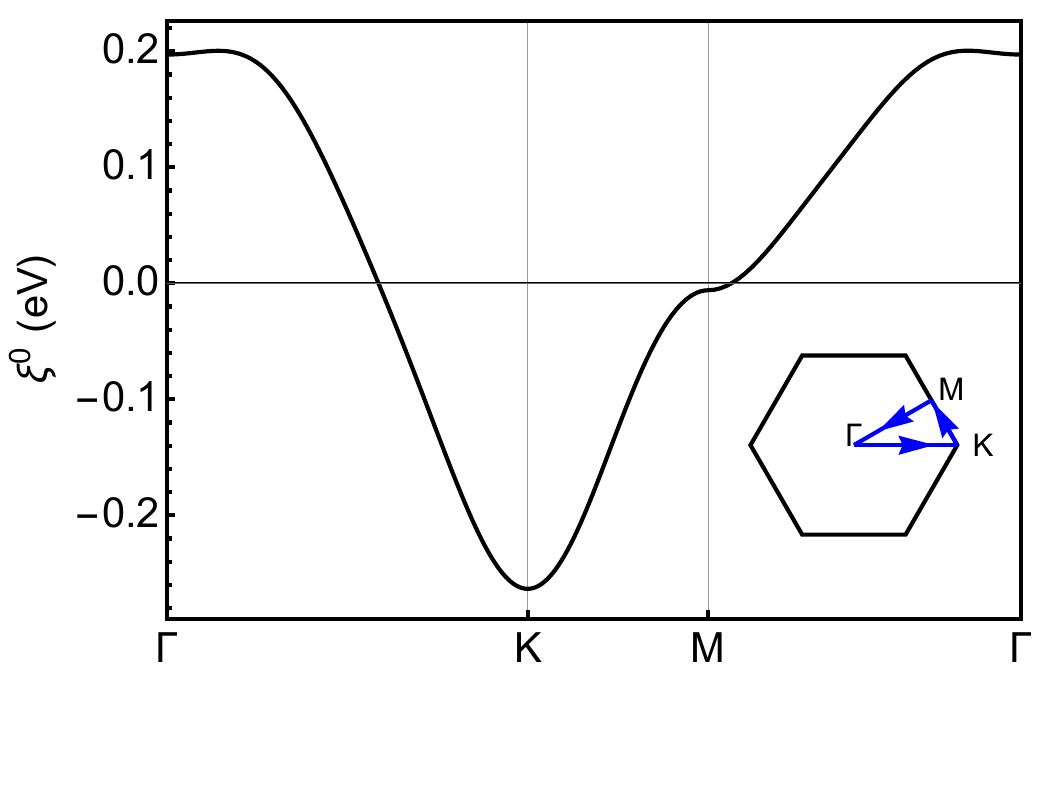}
\caption{
Non-interacting band structure, $\xi^0_{\vec{k}}$.
}
\label{fig:xi_0}
\end{figure}
As the superconductivity in Sn/Si(111) emerges by doping an half-filled AFM Mott insulator,
we model the Hamiltonian of the free bond of the Sn atoms on the $T_4$ sites within the $t-J$ model:
\bea\label{H_tJ}
H=H_{tb}+H_J,
\eea
where:
\begin{subequations}
\bea
H_{tb}=\sum_{ij\sigma}\left(t_{ij}-\mu\delta_{ij}\right)c^\dagger_{i\sigma}c_{j\sigma}
\eea
is the tight binding Hamiltonian, $i,j$ label the sites of the triangular network,
$t_{ij}$ are the hopping amplitudes, $\mu$ is the chemical potential, $c_{i\sigma}$ is the operator for the annihilation of one electron in the site $i$ with spin $\sigma$.
We consider hopping up to the $6^\mathrm{th}$ nearest neighbor,
with the corresponding amplitudes fitted from first principles calculations. We model the additional interaction Hamiltonian, $H_J$, as an AFM exchange term:
\bea\label{H_J}
H_{J}=J\sum_{\langle i,j\rangle}\left(\vec{S}_i\cdot\vec{S}_j-\frac{1}{4}\hat{n}_i\hat{n}_j\right),
\eea
with $J>0$.
The sum runs over all the pairs of nearest neighbor sites,
 $\vec{S}_i=\frac{1}{2}\sum_{\sigma\sigma'}c^\dagger_{i\sigma}\vec{\tau}_{\sigma\sigma'}c_{i\sigma'}$ is the spin operator, $\vec{\tau}=(\tau^x,\tau^y,\tau^z)$ are the Pauli's matrices,
$\hbar=1$ and
$\hat{n}_i=\sum_{\sigma}c^\dagger_{i\sigma}c_{i\sigma}$ is the local number operator.
\end{subequations}
The $t-J$ Hamiltonian \pref{H_tJ} has been intensively studied in the last decades as a model for superconductivity in the high$-T_c$
cuprate superconductors\cite{Anderson_science87,Lee_revmodphys06,Ogata_repprogphys08,Keimer_nature15,Fradkin_revmodphys15,Proust_annrevcm19,Arovas_annrevcm22},
that paradigmatically feature the competition between the Mott insulating and the SC phase. Indeed, besides providing a natural model for AFM ordering it can also describe unconventional superconductivity emerging out of repulsive electronic interactions.
Recently, extended $t-J$ models have been exploited to describe topological SC states in triangular lattices\cite{Jiang_prl20,Huang_prx22}.

\begin{figure}
\centering
\includegraphics[width=\columnwidth]{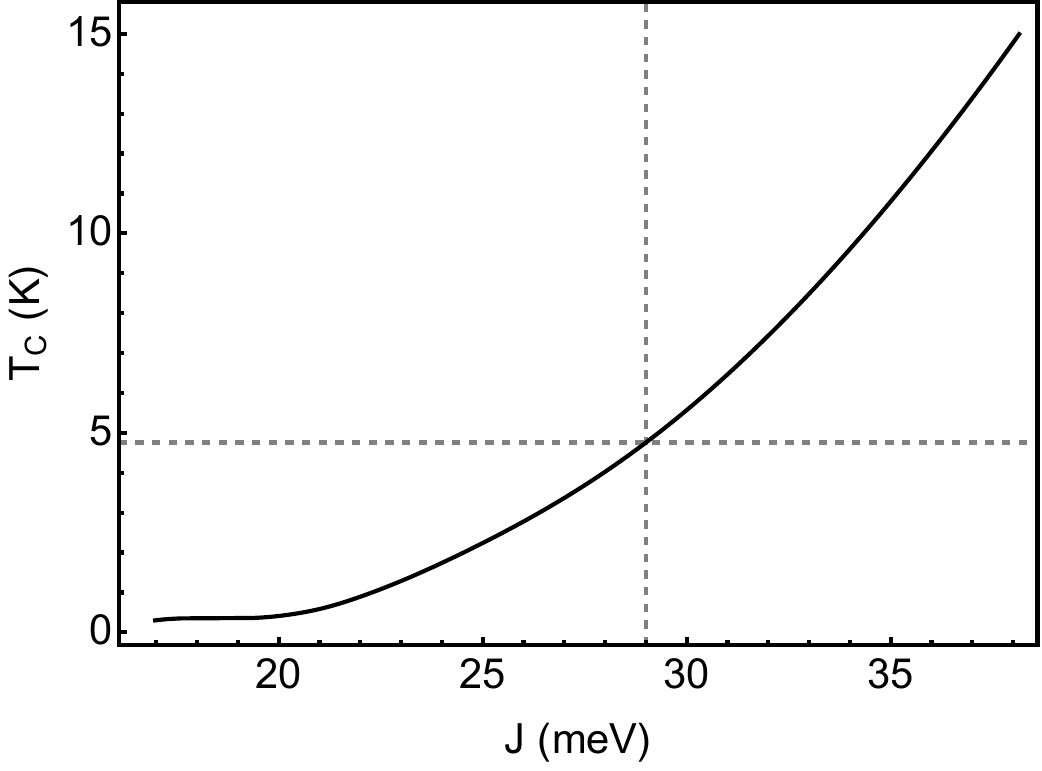}
\caption{
$T_c$ as a function of $J$, obtained for hole-doping: $\delta=10\%$.
The dashed lines identify the value: $J=29$meV, which gives $T_c\simeq4.8$K.
}
\label{fig:Tc_vs_J}
\end{figure}
\begin{figure*}[htb]
\centering
\includegraphics[width=\columnwidth]{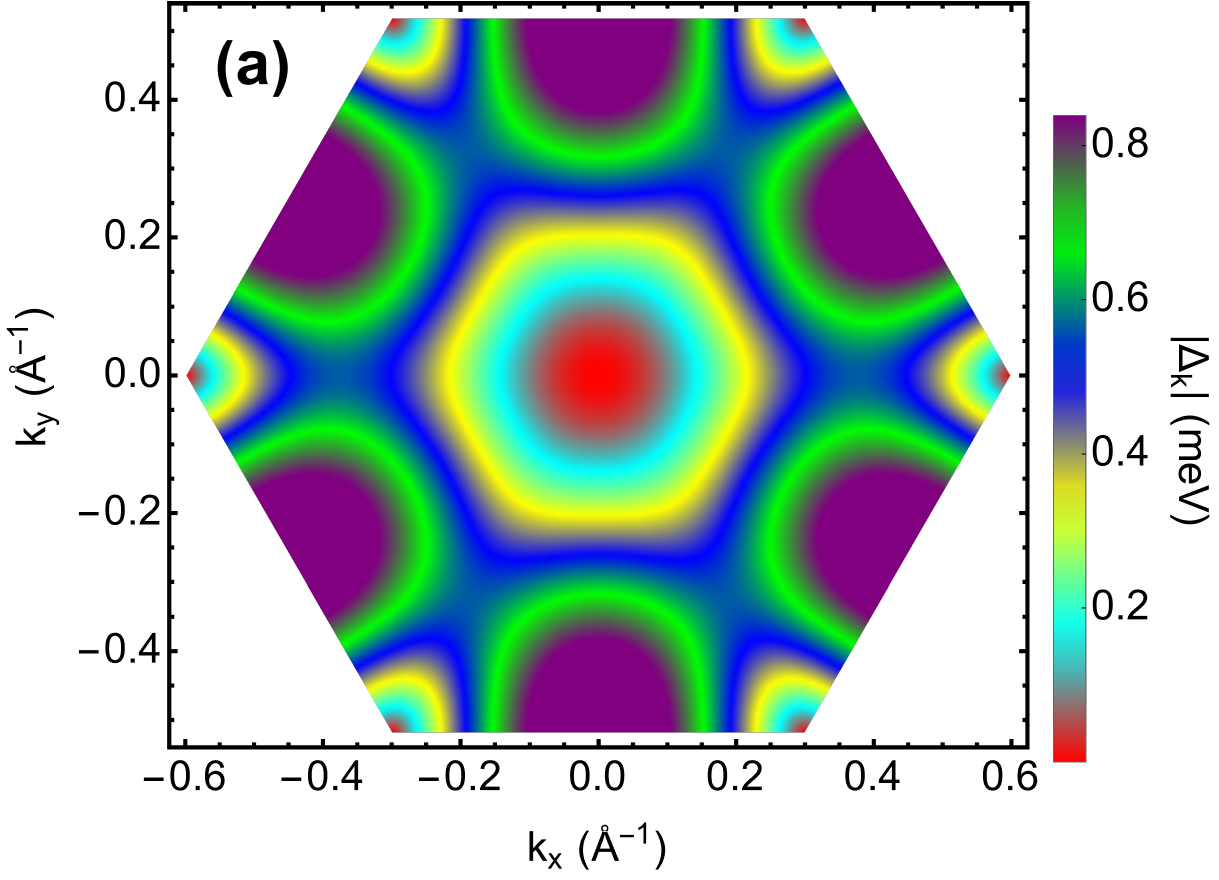}
\includegraphics[width=\columnwidth]{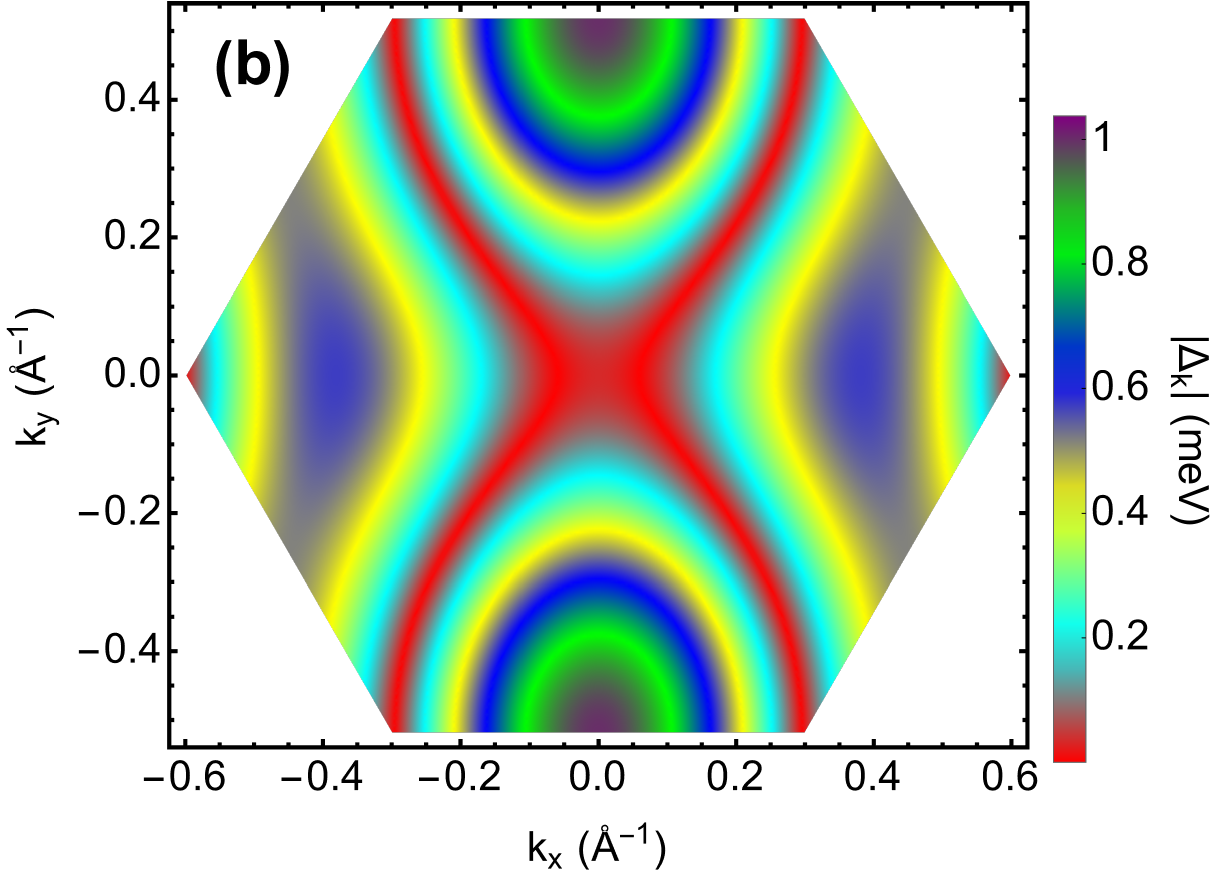}
\caption{
Maps of $|\Delta_{\vec{k}}|$ at $T=0$K, for the solution with $d_{x^2-y^2}+id_{xy}$-wave symmetry(a)
and with $d_{x^2-y^2}$-wave(b).
}
\label{fig:Delta}
\end{figure*}
In the absence of the electromagnetic perturbation, we perform the meanfield approximation to $H$
in both the particle-hole and the particle-particle (pairing) channels, as detailed in the supplementary information \cite{SI}.
This procedure leads to the following quadratic Hamiltonian, expressed in the reciprocal space:
\bea\label{H_MF}
H^{MF}_{tJ}=\sum_{\vec{k}}\left[
\sum_{\sigma}\xi_{\vec{k}}c^\dagger_{\vec{k}\sigma}c_{\vec{k}\sigma}+
\left(\Delta_{\vec{k}}c^{\dagger}_{\vec{k}\uparrow}c^{\dagger}_{-\vec{k}\downarrow}+h.c.\right)
\right]+E_0,
\eea 
where $\vec{k}$ is the wave vector in the first Brillouin zone (BZ),
$\xi_{\vec{k}}=\xi^0_{\vec{k}}+\delta\xi_{\vec{k}}$,
$\xi^0_{\vec{k}}$ is the non-interacting band dispersion computed from the chemical potential,
$\delta\xi_{\vec{k}}$ is the meanfield correction in the particle-hole channel,
$\Delta_{\vec{k}}$ is the pairing amplitude and $E_0$ is an energy constant\cite{SI}.
Both $\delta\xi_{\vec{k}}$ and $\Delta_{\vec{k}}$ can be expressed in terms of the form factors
corresponding to the $s,d_{x^2-y^2},d_{xy}$ symmetries, which are the only ones allowed by the nearest neighbors AFM interaction of the Eq. \pref{H_J}:
$\delta\xi_{\vec{k}}=\sum_\alpha\delta t_\alpha \gamma_\alpha(\vec{k})$, 
$\Delta_{\vec{k}}=\sum_\alpha\Delta_\alpha \gamma_\alpha(\vec{k})$,
where
$\gamma_s(\vec{k})=\left[
\cos(k_xa)+2\cos(k_xa/2)\cos(k_y\sqrt{3}a/2)
\right]/\sqrt{3}$,
$\gamma_{d_{x^2-y^2}}(\vec{k})=\left[
\cos(k_xa)-\cos(k_xa/2)\cos(k_y\sqrt{3}a/2)
\right]\sqrt{2/3}$,
$\gamma_{d_{xy}}(\vec{k})=\sqrt{2}\sin(k_xa/2)\sin(k_y\sqrt{3}a/2)$,
$a=7${\AA} is the lattice constant and $\delta t_\alpha$, $\Delta_\alpha$ are the self-consistent solutions of (see \cite{SI}):
\begin{subequations}
\bea
\delta t_\alpha&=&\frac{J}{2N}\sum_{\vec{k}}\gamma_\alpha(\vec{k})\xi_{\vec{k}}\tanh\left(E_{\vec{k}}/2T\right)/E_{\vec{k}}\\
\Delta_\alpha&=&\frac{J}{N}\sum_{\vec{k}}\gamma_\alpha(\vec{k})\Delta_{\vec{k}}\tanh\left(E_{\vec{k}}/2T\right)/E_{\vec{k}},
\eea
\end{subequations}
where $N$ is the number of lattice sites,
$E_{\vec{k}}=\sqrt{\xi^2_{\vec{k}}+|\Delta_{\vec{k}}|^2}$,
$T$ is the temperature and we set $k_B=1$.
The non-interacting band structure, $\xi^0_{\vec{k}}$, is shown in the Fig. \ref{fig:xi_0}
along the high symmetry path of the BZ specified in the inset panel.
Having fixed $\xi^0_{\vec{k}}$, $T_c$ is determined by the interaction strength $J$.
In order to reproduce the experimental results of the Ref. \cite{Wu_prl20}, we fix the value of $J$
by imposing: $T_c=4.8$K and hole-doping: $\delta=10\%$.
This is illustrated by the Fig. \ref{fig:Tc_vs_J}, showing $T_c$ as a function of $J$. As outlined by the dashed lines, we finally obtain: $J=29$meV, wich is of the same order of magnitude as that obtained by first principles calculations ($J_{fp}=10$meV, see also \cite{SI}).
We find two distinct SC solutions\cite{SI}, one featuring chiral $d_{x^2-y^2}+id_{xy}$-wave symmetry, as claimed in the Ref. \cite{Ming_nature23},
and the other with pure $d$-wave symmetry. 
The maps of $|\Delta_{\vec{k}}|$ at $T=0K$ are shown in the Fig. \ref{fig:Delta}(a)-(b)
for the two cases, emphasizing that the $C_6$ symmetry is preserved by the chiral solution,
while it is broken by the pure $d$-wave, which only features $C_2$.
It's worth noting that, for the pure $d$-wave case, the system is symmetric under continuum $U(1)$ rotations in the $\Delta_{d_{x^2-y^2}}-\Delta_{d_{xy}}$ plane. Therefore, we have chosen the initial conditions of the meanfield calculation in order to select the pure $d_{x^2-y^2}$-wave symmetry, which is the one shown in the Fig. \ref{fig:Delta}(b).
As we show in \cite{SI}, the chiral symmetry is energetically favored by a very small amount: $\sim10^{-7}$eV per site, allowing us to safely consider the two solutions as almost degenerate.

\section{Tracking the evolution of the GS in a time-dependent electric field}
We introduce a uniform vector potential, $\vec{A}(t)$, in the Hamiltonian \pref{H_MF}
via the Peierls substitution: $\xi^0_{\vec{k}}\rightarrow \xi^0_{\vec{k}-e\vec{A}(t)}$,
where $e$ is the electron charge and $c=1$.
It is convenient to describe the dynamics of the meanfield GS in terms of the quantum average of the Anderson's pseudo-spin:
$\vec{\sigma}_{\vec{k}}=\left\langle\Psi^\dagger_{\vec{k}}\vec{\tau}\Psi_{\vec{k}}\right\rangle$, where:
$\Psi_{\vec{k}}=\left(c_{\vec{k}\uparrow},c^\dagger_{-\vec{k}\downarrow}\right)^T$.
At the meanfield level, $\vec{\sigma}_{\vec{k}}$ describes a precession motion given by the equations:
\bea\label{Anderson_precession}
\frac{\,d \vec{\sigma}_{\vec{k}}}{\,dt}(t)=2\vec{b}_{\vec{k}}(t)\times \vec{\sigma}_{\vec{k}}(t),
\eea
with: $\vec{b}_{\vec{k}}=\left(\Delta'_{\vec{k}},-\Delta''_{\vec{k}},\xi_{\vec{k}}\right)$.
Imposing:
$\Delta'_\alpha=-\frac{J}{N}\sum_{\vec{k}}\gamma_\alpha(\vec{k})\sigma^x_{\vec{k}}$,
$\Delta''_\alpha=\frac{J}{N}\sum_{\vec{k}}\gamma_\alpha(\vec{k})\sigma^y_{\vec{k}}$ and
$\delta t_\alpha=-\frac{J}{2N}\sum_{\vec{k}}\gamma_\alpha(\vec{k})\sigma^z_{\vec{k}}$,
one has to solve the Eqs. \pref{Anderson_precession} self-consistently at any time,
which is equivalent to solving the Bloch's equations.
In this way one can track the dynamics induced by the incident field on both the gap's function, $\Delta_{\vec{k}}(t)$,
and the occupation number, $n_{\vec{k}}(t)$, including all the non-linear effects emerging for very intense fields.
While $\Delta_{\vec{k}}(t)$ describes the evolution of the SC condensate, $n_{\vec{k}}(t)$ defines the electromagnetic current flowing through the sample, according to:
\be
\label{eqj}
\vec{J}(t)=-\sum_{\vec{k}}\nabla_{\vec{A}}\xi^0_{\vec{k}-e\vec{A}(t)}n_{\vec{k}}(t).
\ee
Experimentally, $\vec{J}$ can be measured either directly by the voltage difference at the edge of the sample, or indirectly by the electric field reflected by the sample. Transmission measurements are not allowed because of the Si substrate.

\section{Numerical results: the angle-resolved THG}
\begin{figure}
\centering
\includegraphics[width=\columnwidth]{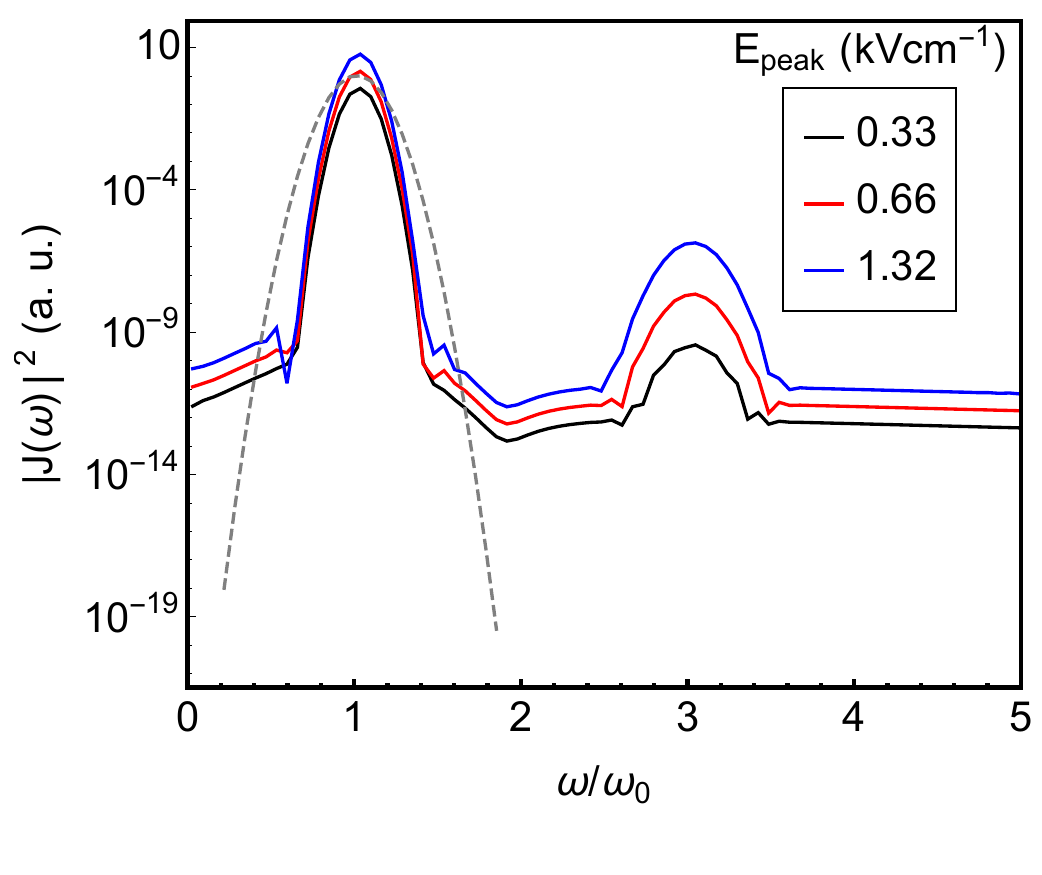}
\caption{
Power spectrum as a function of the frequency,
obtained in the chiral solution, at $T=0$, for three different values of $E_{peak}$.
The dashed gray line shows the power spectrum of the incident field, $|\vec{E}(\omega)|^2$.
The incident field is polarized along one of the crystallografic direction.
}
\label{fig:PS}
\end{figure}

The current \pref{eqj} contains explicitly the response of the system at all orders in ${\bf A}$, since the vector potential enters both in the velocity term (gradient of the band dispersion) and in the computation of the time evolution of $n_{\bf k}(t)$. For weak field one recovers the usual linear-response result, $J^L\sim \chi^{(1)} A$, that for a purely monochromatic field at frequency $\omega_0$ implies an induced current at the same frequency $\omega_0$. On the other hand, for increasing strength of the driving field one can generate a non-linear current whose lowest order scales as $J^{NL}\sim \chi^{(3)} A^3$, so that in the induced current one observes higher harmonics of the driving field, the lowest one being at $3\omega_0$. Since the THG measurements are usually performed with multicycle light pulses\cite{Matsunaga_science14,Matsunaga_prb17,Chu_natcomm20,shimano_commphys21,wang_mgb2_prb21,wang_mgb2_prb22,kaiser_natcomm23,kaiser_cm23}, that are not perfectly monochromatic, in the numerical calculations we model the vector potential as an oscillatory function at the central frequency $\omega_0$ convoluted with gaussian envelope:
$\vec{A}(t)=A_0e^{-(t/\tau)^2}\sin(\omega_0 t)\vec{n}$,
where $A_0$ defines the intensity of the field, $\tau$ the duration of the pulse,
and $\vec{n}$ is the polarization vector, with $|\vec{n}|=1$. Since we are interested to exploit a possible resonance of $\omega_0$ with the gap value, we choose $\omega_0=1$meV, which is of the order of $\Delta_{\vec{k}}$, and $\tau=11$ps, so that the frequency resolution of the pulse is $\Delta \omega/\omega_0\simeq 0.5$.
The electric field is given by: $\vec{E}(t)=-\partial_t \vec{A}(t)$.
The Fig. \ref{fig:PS} shows the power spectrum, $|\vec{J}(\omega)|^2$,
as a function of the frequency, obtained in the chiral solution, at $T=0$, for three different intensities of the incident field,
as specified in the legend, $E_{peak}$ being the maximum of $|\vec{E}(t)|$.
Here the incident field is polarized along one of the crystallografic direction, the $x$ axis by convention.
The dashed gray line shows the power spectrum of the incident field, $|\vec{E}(\omega)|^2$.
As expected, the intensity of the THG at $\omega=3\omega_0$ increases with $E_{peak}$
much faster than the first harmonic, the latter being dominated mostly by the linear response.

\begin{figure*}
\centering
\includegraphics[width=\textwidth]{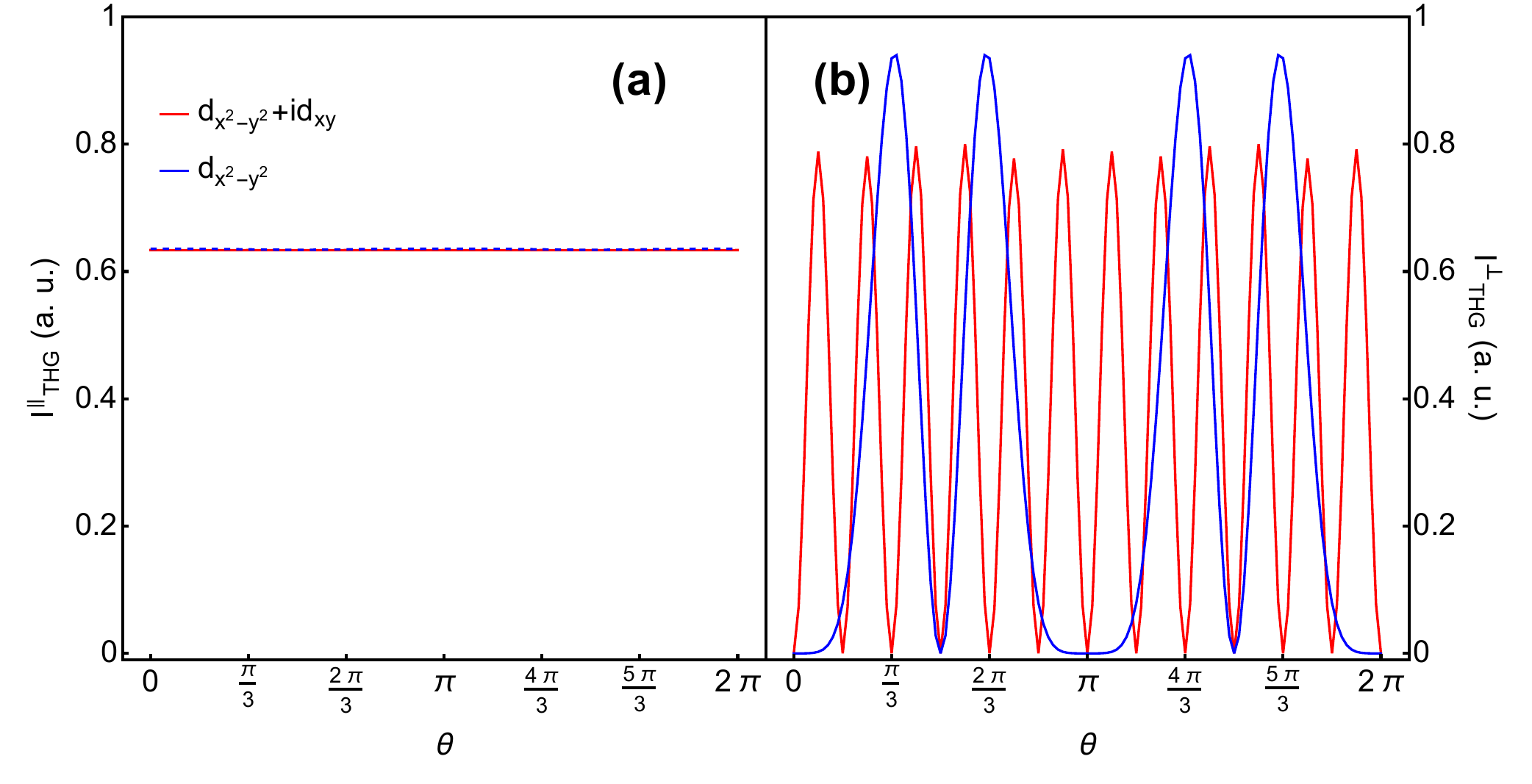}
\caption{
$I^{\|}_{THG}$(a) and $I^{\perp}_{THG}$(b)
as a function of $\theta$,
corresponding to $E_{peak}=0.66$kVcm$^{-1}$ and obtained for the chiral $d_{x^2-y^2}+id_{xy}$-wave solution (red line) and for the pure
$d_{x^2-y^2}$-wave (blue line).
}
\label{fig:I_THG}
\end{figure*}

Next, we study the THG by varying
the polarization of the incident field: $\vec{n}=\cos\theta\hat{x}+\sin\theta\hat{y}$.
In particular, we focus on the two quantities: $I^{\|}_{THG}=\left|\vec{n}\cdot\vec{J}(3\omega_0)\right|$
and $I^{\perp}_{THG}=\left|\vec{n}^{\perp}\cdot\vec{J}(3\omega_0)\right|$,
representing the intensity of the THG in the direction parallel and orthogonal to the incident field, respectively.
The Fig. \ref{fig:I_THG} shows $I^{\|}_{THG}$(a) and $I^{\perp}_{THG}$(b)
as a function of $\theta$, corresponding to $E_{peak}=0.66$kVcm$^{-1}$ and obtained for the chiral $d_{x^2-y^2}+id_{xy}$-wave solution (red line) and for the pure
$d_{x^2-y^2}$-wave (blue line).
As is evident, while $I^{\|}_{THG}$ is basically constant,
$I^{\perp}_{THG}$ displays typical oscillations.
We expect that the isotropic behavior of $I^{\|}_{THG}$ is mainly due to the leading contribution
of the instantaneous electronic response, which is polarization-independent\cite{Cea_prb18}.
If we then focus on $I^{\perp}_{THG}$, we note a marked difference between the two gap's symmetries:
in the chiral symmetry, $I^{\perp}_{THG}$ has a period of $\pi/3$, following from the $C_6$ symmetry of the crystal;
in contrast, in the pure $d$-wave symmetry, the period is $\pi$, as a consequence of the $C_3$ symmetry breaking which preserves only $C_2$.
Therefore, the behavior of the THG as a function of the polarization of the incident field
provides an experimental tool to identify the symmetry of the SC gap in the boron-doped Sn/Si(111),
distinguishing between the chiral $d_{x^2-y^2}+id_{xy}$- and the pure $d$-wave.
It's worth noting that the periodicity of $\pi/3$ cannot in principle distinguish the chiral $d_{x^2-y^2}+id_{xy}$- from the $s$-wave, which preserves $C_6$ too.
However, the $s$-wave can be easily ruled out by the STM spectra reported in the Ref. \cite{Ming_nature23}.
Finally, in the experiments the polarization dependence of $I_{THG}$ might be sensitive to the effects of disorder\cite{seibold_prb21,Benfatto_prb23}, not considered here, challenging the use of ultra-clean samples. Indeed, when disorder is present paramagnetic-like processes mediating the coupling of the light to SC excitations become possible, providing additional channels to generate a non-linear response\cite{silaev_prb19,tsuji_prr20,seibold_prb21,Benfatto_prb23}. So far numerical studies for square lattices have shown that these additional contributions triggered by disorder tend to soften the polarization dependence of the $I^\parallel_{THG}$ signal, irrespective of the pairing symmetry\cite{seibold_prb21,Udina_fardisc22,Benfatto_prb23}. On the other hand, the perpendicular component is expected to still retain the angular dependence of the clean case, at least in the limit where paramagnetic effects are not quantitatively dominant. On this respect, the present calculations performed in the clean limit should remain valid for what concerns the dependence of $I^\perp_{THG}$ on the gap symmetry even when one includes  realistic disorder effects, that are beyond the scope of the present manuscript.

\section{Conclusions}

We study the superconductivity reported recently in the Sn/Si(111)
ad-layers doped with boron.
Assuming that the interactions responsible for the pairing are of AFM nature, we describe the system by the $t-J$ model.
Our meanfield analysis well reproduces the value of the AFM coupling, $J$, predicted by first principle calculations and the $d$-wave symmetry of the SC OP claimed by recent experimental findings\cite{Ming_nature23}.
By perturbing the system with an intense THz electromagnetic field, we focus on the leading non-linear effects, the THG, showing that its dependence on the polarization of the incident field can distinguish between the chiral $d_{x^2-y^2}+id_{xy}$- and the pure $d$-wave symmetry of the SC gap.
Therefore, our work i) provides a minimal efficient theoretical model for describing the pairing in the
emerging superconductor Sn/Si(111) and
ii) paves the way for unveiling the pairing symmetry therein
by means of the THz pump protocol, a method that has succeeded so far in investigating both conventional and high-$T_c$ cuprate superconductors.

\section*{Acknowledgements}
We thank Andrea Marini for helpful and interesting conversations.

T. C., C. T. and G. P. acknowledge support from CINECA for computational resources through ISCRA projects.
Research at SPIN-CNR has been funded by the European Union - NextGenerationEU under the Italian Ministry of University and Research (MUR) National Innovation Ecosystem grant ECS00000041 - VITALITY, C. T. acknowledges Università degli Studi di Perugia and MUR for support within the project Vitality.
C. T. acknowledges financial support from the Italian Ministry for Research and Education through PRIN-2022 project ``DARk-mattEr-DEVIces-for-Low-energy-detection - DAREDEVIL'' (IT-MIUR Grant No. 2022Z4RARB). L.B. acknowledges financial support by Sapienza University under projects Ateneo (RM12117A4A7FD11B and RP1221816662A977) and by ICSC–Centro Nazionale di Ricerca in High Performance Computing, Big Data and Quantum Computing, funded by European Union – NextGenerationEU.

G. P. acknowledges financial support from the Italian Ministry for Research and Education through PRIN-2017 project ``Tuning and understanding Quantum phases in 2D materials - Quantum 2D" (IT-MIUR Grant No. 2017Z8TS5B) and fundings from the European Union - NextGenerationEU under the Italian Ministry of University and Research (MUR) National Innovation Ecosystem grant ECS00000041 - VITALITY - CUP E13C22001060006.

\bibliography{Literature}


\clearpage

\onecolumngrid

\setcounter{section}{0}
\setcounter{equation}{0}
\setcounter{figure}{0}
\setcounter{table}{0}
\setcounter{page}{1}
\makeatletter
\renewcommand{\theequation}{S\arabic{equation}}
\renewcommand{\thefigure}{S\arabic{figure}}

\begin{center}
\Large Supplementary information for:\\
\textbf{Unveiling the pairing Symmetry of the superconducting Sn/Si(111) via angle-resolved THz pump spectroscopy}
\end{center}

\section{The meanfield approximation for the $t-J$ model}
We first consider the tight binding Hamiltonian:
\bea
H_{tb}=\sum_{ij\sigma}\left(t_{ij}-\mu\delta_{ij}\right)c^\dagger_{i\sigma}c_{j\sigma}=
\sum_{\vec{k}\sigma}\xi^0_{\vec{k}}c^\dagger_{\vec{k}\sigma}c_{\vec{k}\sigma},
\eea
where:
\bea
\xi^0_{\vec{k}}=\sum_it_{ij}e^{-i\vec{k}\cdot\left(\vec{R}_i-\vec{R}_j \right)}-\mu,
\eea
is the non-interacting band dispersion computed from $\mu$ and the $\vec{R}_i$'s are the lattice coordinates.
In the numerical calculations, we compute $\xi^0_{\vec{k}}$ by truncating the sum over the coordinates at the 6$^\mathrm{th}$ nearest neighbor.

\begin{figure}
\centering
\includegraphics[width=\columnwidth]{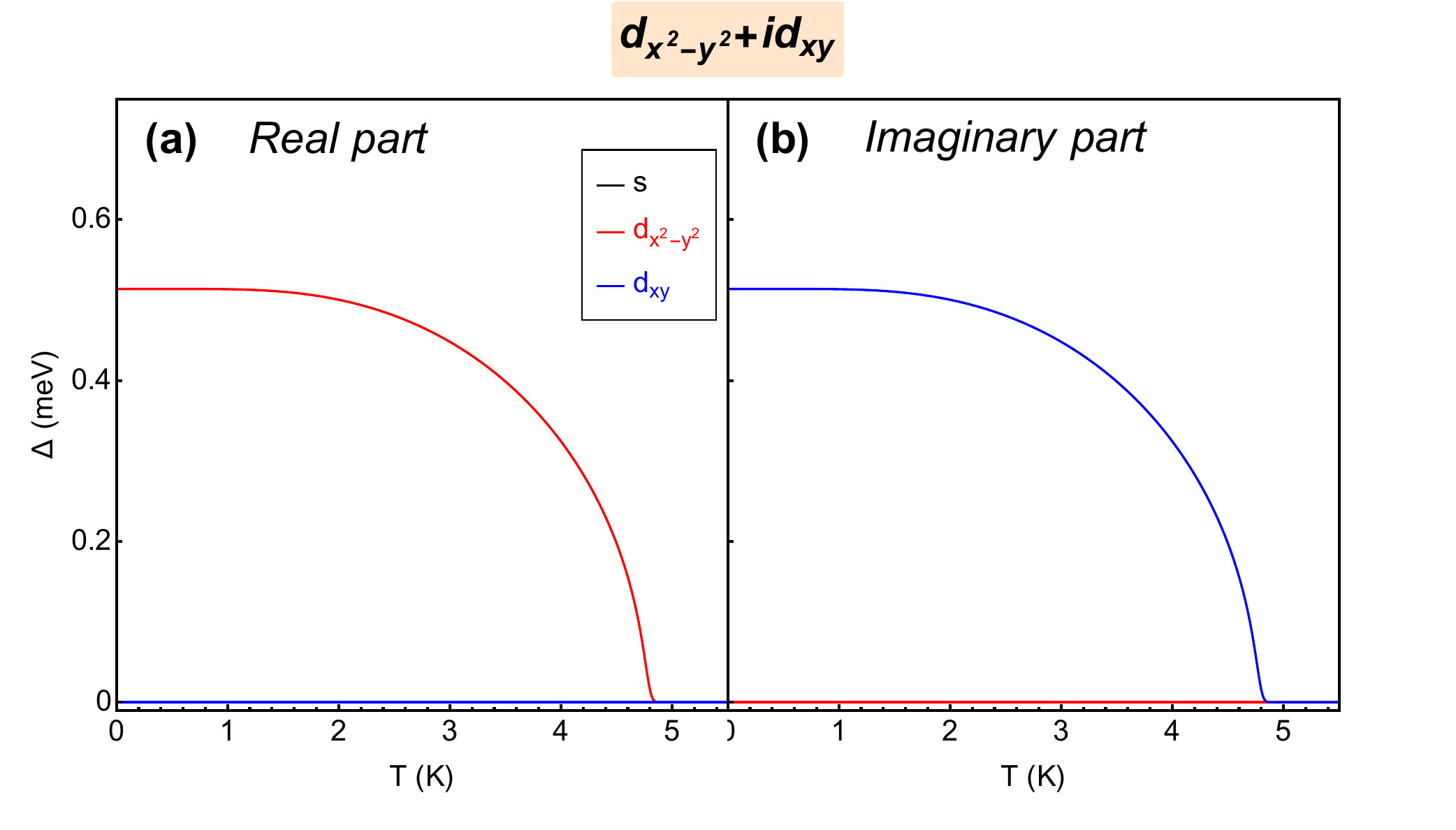}
\includegraphics[width=\columnwidth]{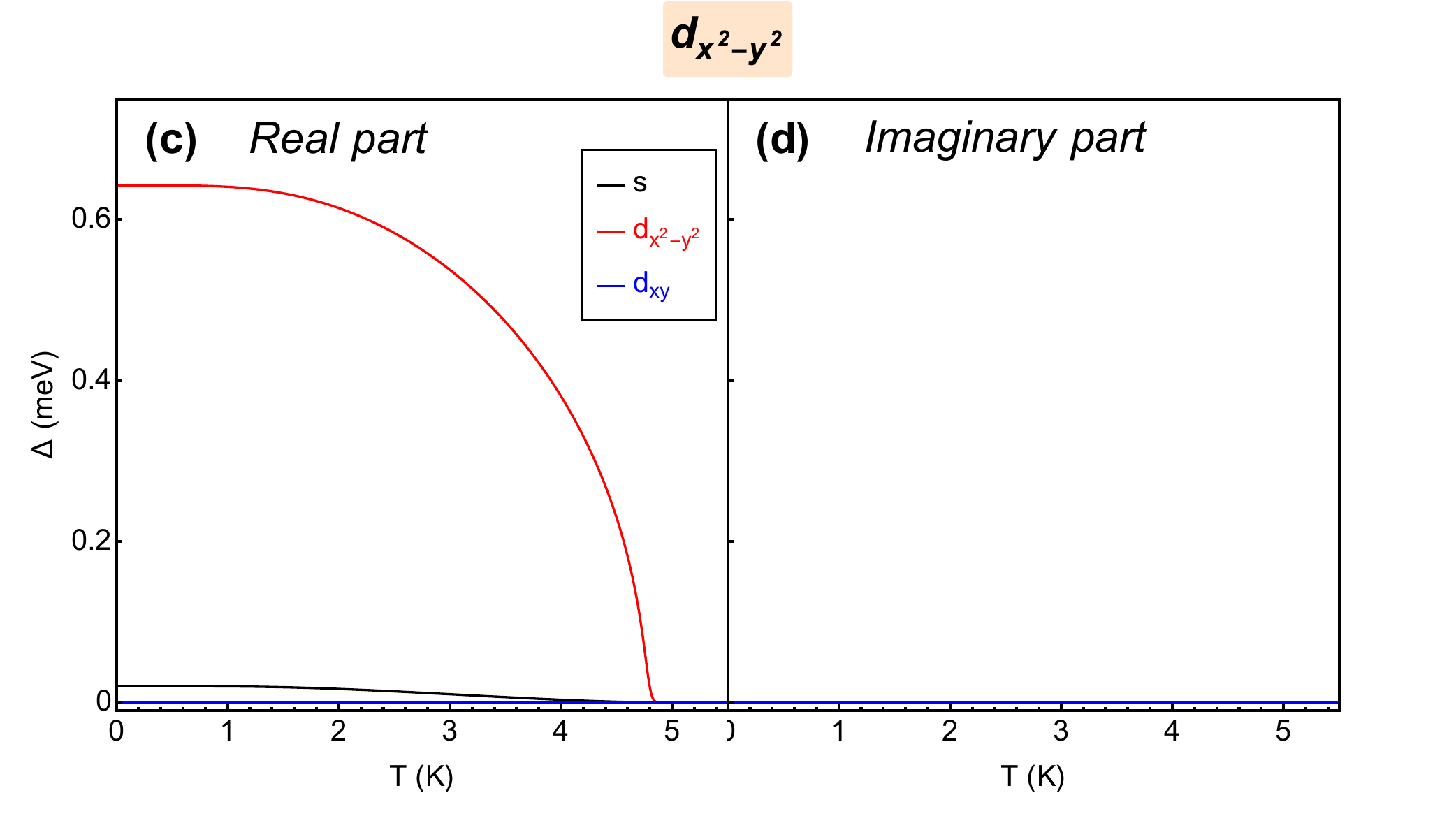}
\caption{
$\Delta_\alpha$ as a function of the temperature, obtained for the chiral $d_{x^2-y^2}+id_{xy}$,solution (a),(b) and for the pure $d$ one (c),(d). The panels (a),(c) refer to the real part of $\Delta_\alpha$, while the panels (b),(d) to the imaginary part. The different symmetries are color coded as specified in the inset panel.
}
\label{fig:Delta_vs_T}
\end{figure}

Next, we consider the AFM Hamiltonian:
\bea
H_{J}=J\sum_{\langle i,j\rangle}\left(\vec{S}_i\cdot\vec{S}_j-\frac{1}{4}\hat{n}_i\hat{n}_j\right).
\eea
In the momentum space, it can be written as:
\bea\label{H_J_q}
H_{J}=\frac{J}{N}\sum_{\vec{q}}\sum_{\alpha=1}^3\cos(\vec{q}\cdot\vec{a}_\alpha)
\left(\vec{S}_{\vec{q}}\cdot\vec{S}_{-\vec{q}}-\frac{1}{4}\hat{n}_{\vec{q}}\hat{n}_{-\vec{q}}\right),
\eea
where:
\begin{subequations}
\bea
\vec{S}_{\vec{q}}&=&\frac{1}{2}\sum_{\vec{k}\sigma\sigma'}c^\dagger_{\vec{k}-\vec{q}/2,\sigma}
\vec{\tau}_{\sigma\sigma'}c_{\vec{k}+\vec{q}/2,\sigma'},\\
\hat{n}_{\vec{q}}&=&\sum_{\vec{k}\sigma}
c^\dagger_{\vec{k}-\vec{q}/2,\sigma}
c_{\vec{k}+\vec{q}/2,\sigma}
\eea
\end{subequations}
and $\vec{a}_1=a(1,0),\vec{a}_2=a\left(-1/2,\sqrt{3}/2\right),\vec{a}_3=a\left(-1/2,-\sqrt{3}/2\right)$ are unit lattice vectors.
After a lengthy calculations, one can easily rewrite the Eq. \pref{H_J_q} as:
\bea
H_{J}=\frac{J}{N}\sum_{\vec{q}\vec{k}\vec{p}}
\sum_{\alpha=1}^3\cos(\vec{q}\cdot\vec{a}_\alpha)
\left(
c^\dagger_{\vec{k}-\vec{q}/2,\uparrow}
c_{\vec{k}+\vec{q}/2,\downarrow}
c^\dagger_{\vec{p}+\vec{q}/2,\downarrow}
c_{\vec{p}-\vec{q}/2,\uparrow}-
c^\dagger_{\vec{k}-\vec{q}/2,\uparrow}
c_{\vec{k}+\vec{q}/2,\uparrow}
c^\dagger_{\vec{p}+\vec{q}/2,\downarrow}
c_{\vec{p}-\vec{q}/2,\downarrow}
\right).
\eea
To perform the meanfield approximation of $H_J$,
we consider all the quantum averages preserving spin and momentum, 
we neglect the quantum fluctuations and we require the time-reversal symmetry.
This yields:
\bea\label{suppl:H_MF_J}
H^{MF}_{J}=
-\frac{J}{N}\sum_{\vec{k}\vec{p}}\sum_{\alpha=1}^3
\gamma_\alpha(\vec{k})\gamma_\alpha(\vec{p})\left[
2\left(\left\langle c_{-\vec{p}\downarrow}c_{\vec{p}\uparrow}\right\rangle
c^\dagger_{\vec{k}\uparrow}c^\dagger_{-\vec{k}\downarrow}+h.c. \right)+
\frac{n_{\vec{p}}}{2}\sum_\sigma c^\dagger_{\vec{k}\sigma}c_{\vec{k}\sigma}
\right]+E_0,
\eea
where:
\bea
E_0=\frac{2J}{N}\sum_{\alpha}\left|\sum_{\vec{k}}\gamma_\alpha(\vec{k})\left\langle
c_{-\vec{k}\downarrow}c_{\vec{k}\uparrow}
\right\rangle\right|^2+
\frac{J}{4N}
\sum_{\alpha}\left[\sum_{\vec{k}}\gamma_\alpha(\vec{k})
n_{\vec{k}}
\right]^2
\eea
is an energy constant and
$n_{\vec{k}}=\sum_{\sigma}\left\langle
c^\dagger_{\vec{k}\sigma}c_{\vec{k}\sigma}
\right\rangle$.
Note that we have neglected the terms involving only a constant shift of the chemical potential in the Eq. \pref{suppl:H_MF_J}. The meanfield Hamiltonian is the sum:
\bea
H_{tJ}^{MF}=H_{tb}+H^{MF}_J.
\eea
Now we define:
\begin{subequations}
\bea
\Delta_\alpha&=&-\frac{2J}{N}\sum_{\vec{k}}\gamma_\alpha(\vec{k})\left\langle c_{-\vec{k}\downarrow}c_{\vec{k}\uparrow}\right\rangle\text{ , }
\Delta_{\vec{k}}=\sum_{\alpha}\gamma_\alpha(\vec{k})\Delta_\alpha
\\
\delta t_{\alpha}&=&-\frac{J}{2N}\sum_{\vec{k}}\gamma_\alpha(\vec{k})n_{\vec{k}}\text{ , }
\delta\xi_{\vec{k}}=\sum_{\alpha}\gamma_\alpha(\vec{k})\delta t_{\alpha}\text{ , }
\xi_{\vec{{k}}}=\xi^0_{\vec{{k}}}+\delta\xi_{\vec{k}},
\eea
\end{subequations}
which allows to write $H_{tJ}^{MF}$ as:
\bea\label{suppl:H_MF_tJ}
H^{MF}_{tJ}=\sum_{\vec{k}}\left[
\sum_{\sigma}\xi_{\vec{k}}c^\dagger_{\vec{k}\sigma}c_{\vec{k}\sigma}+
\left(\Delta_{\vec{k}}c^{\dagger}_{\vec{k}\uparrow}c^{\dagger}_{-\vec{k}\downarrow}+h.c.\right)
\right]+E_0,
\eea
and:
\bea
E_0=\frac{N}{J}\sum_\alpha\left(\frac{\left|\Delta_\alpha\right|^2}{2}+\delta t^2_\alpha\right).
\eea
$\Delta_\alpha$ and $\delta t_\alpha$ have to be obtained self-consistently by using the Hamiltonian \pref{suppl:H_MF_tJ} to compute the quantum averages. This yields the self-consistent equations:
\begin{subequations}
\bea
\Delta_\alpha&=&\frac{J}{N}\sum_{\vec{k}}\gamma_\alpha(\vec{k})\frac{\Delta_{\vec{k}}}{E_{\vec{k}}}\tanh\left(E_{\vec{k}}/2T\right),\\
\delta t_\alpha&=&\frac{J}{2N}\sum_{\vec{k}}\gamma_\alpha(\vec{k})
\frac{\xi_{\vec{k}}}{E_{\vec{k}}}\tanh\left(E_{\vec{k}}/2T\right),
\eea 
\end{subequations}
that must be solved along with the equation for the electronic density:
\bea
n=1-\frac{1}{N}\sum_{\vec{k}}\frac{\xi_{\vec{k}}}{E_{\vec{k}}}\tanh\left(E_{\vec{k}}/2T\right),
\eea
which sets the value of $\mu$. In this work we consider: $n=0.9$, corresponding to the hole doping: $\delta=10\%$.
The Fig. \ref{fig:Delta_vs_T} shows $\Delta_\alpha$ as a function of the temperature, obtained for the chiral $d_{x^2-y^2}+id_{xy}$,solution (a),(b) and for the pure $d$ one (c),(d). The panels (a),(c) refer to the real part of $\Delta_\alpha$, while the panels (b),(d) to the imaginary part. The different symmetries are color coded as specified in the inset panel.

\begin{figure}
\centering
\includegraphics[width=\columnwidth]{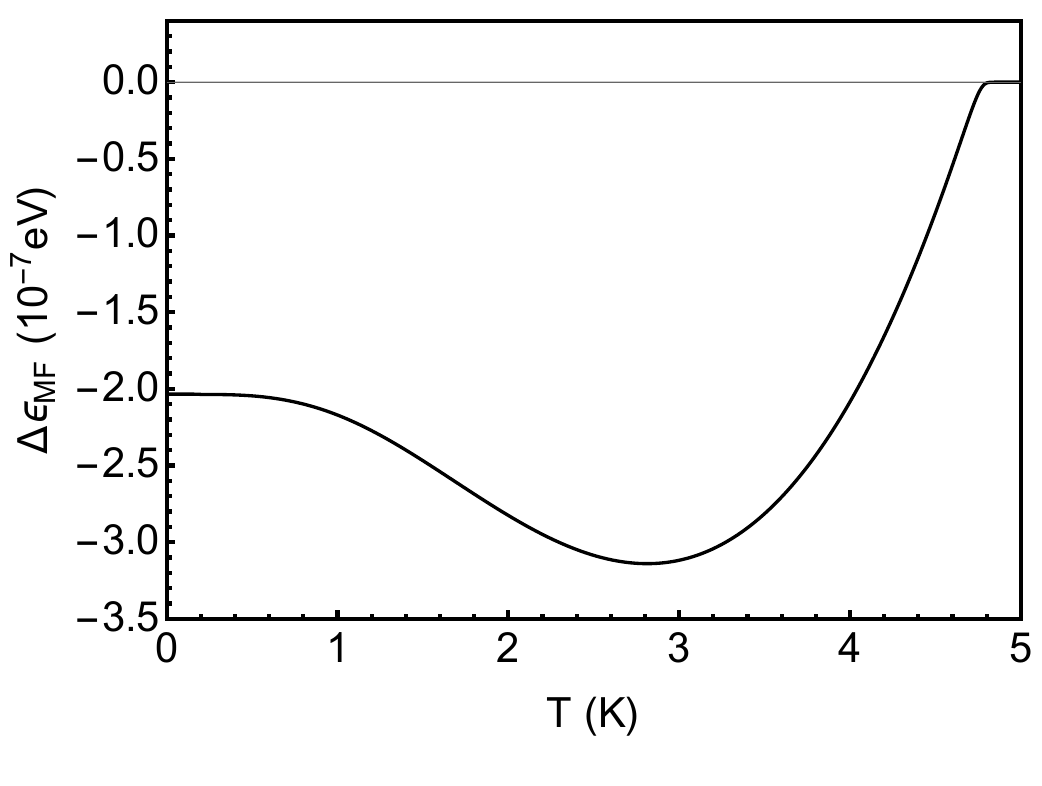}
\caption{
Difference between the meanfield energy per site of the chiral solution and that of the non-chiral one, as a function of the temperature.
}
\label{fig:Delta_e_MF}
\end{figure}

The meanfield energy per site is given by:
\bea
\epsilon_{MF}=\frac{E_0}{N}+\frac{1}{N}\sum_{\vec{k}}\left[
\xi_{\vec{k}}-E_{\vec{k}}\tanh\left(E_{\vec{k}}/2T\right)
\right]+\mu n,
\eea
where the first two terms come out from averaging $H^{MF}_{tJ}$, while the last term has to be considered in $\epsilon_{MF}$ because the Hamiltonian $H_{tJ}$ is defined in the grancanonical ensemble.
The Fig. \ref{fig:Delta_e_MF} shows the difference, $\Delta\epsilon_{MF}$, between the meanfield energy per site of the chiral solution and that of the non-chiral one, as a function of the temperature.

\newpage
\section{Evaluation of the AFM coupling from first principles calculations}
We evaluate the nearest-neighbor exchange constant ($J$) between different Sn atoms through the Hubbard model described by the following Hamiltonian:

\bea
H=J_{fp}\sum_{\langle i, j \rangle} \vec{S}_i\cdot\vec{S}_j
\eea

We adopted a super-cell approach considering a 2$\sqrt{3}\times3$ cell, with respect to the Si(111) surface periodicity, with different collinear
magnetic configurations: two Sn atoms per cell magnetized ferromagnetic or anti-ferromagnetic each other. 

Accordingly with literature\cite{Profeta_prl2007,Tresca_prl18,Tresca_prb21,Tresca_prb23}, we modeled the $\alpha$-Sn/Si(111) surface by considering a layer of Sn atoms on top of three Si-bilayers; the Si-dangling bonds at the opposite side are capped with hydrogen atoms fixed to the relaxed positions obtained by capping the pristine Si(111) surface at one side. 

The atomic position of the first five atomic layers are optimized (Sn atoms and the first two Si-bi-layers) whereas the remaining bi-layer is fixed to the Si bulk positions. More than 15~\AA\ of vacuum is included.
Density functional theory calculations are performed with the \textsc{Quantum-Espresso}\cite{QEcode,QE-2017} code. We used ultrasoft pseudopotentials with an energy cutoff up to 45~Ry. Integration over the Brillouin zone was performed using uniform 12$\times 6\times 1 $ grid. 

\begin{figure}[h!]
\centering
\includegraphics[width=0.75\columnwidth]{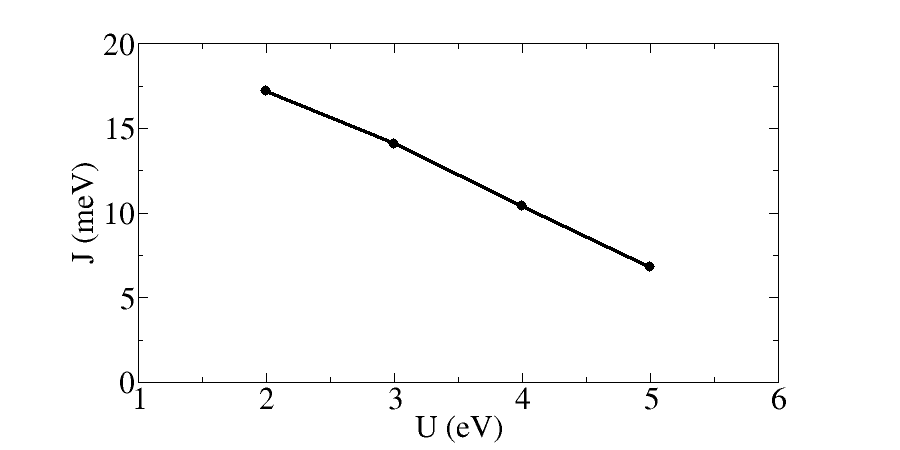}
\caption{
Evolution of the antiferromagnetic exchange coupling as a function of the Hubbard term $U$.
}
\label{fig:S_AFM_J}
\end{figure}

The magnetic calculations were done using the semilocal approximation for the exchange and correlation term in a GGA+U framework\cite{Profeta_prl2007}.
As showh in Fig.\ref{fig:S_AFM_J}, the obtained results appear to be "slightly" affected by the choice of the $U$ parameter: in the range $U=$2\textdiv 5~eV, $J_{fp}$ is monotonically decreasing as a function of $U$, without change of sign or order of magnitude.  

The most favourable magnetic configuration is the anti-ferromagnetic one, with an energy difference of the order of $\Delta E_{FM-AFM}\sim 35$\textdiv 15~meV with respect to the the ferromagnetic phase.
The calculated antiferromagnetic exchange coupling results $J_{fp}\sim10$~meV (for $U=4$~eV, $\Delta E_{FM-AFM}=0.021$~eV)\cite{Profeta_prl2007}.

\end{document}